\documentclass[conference]{IEEEtran}

\usepackage{amsmath}
\usepackage{graphicx}%

\usepackage{units}
\usepackage{url,xspace}
\usepackage{nicefrac}
\usepackage{here}
\usepackage[np]{numprint}
\usepackage{units}
\usepackage{paralist}
\usepackage{hyperref}
\usepackage{booktabs}
\usepackage{paralist}
\usepackage{amsmath,amssymb}

\usepackage{enumitem}

\usepackage{tabularx}

\usepackage[colornames]{xcolor}

\definecolor{ude}{RGB}{0,76,147}
\definecolor{udeBeige}{RGB}{239,228,191}
\definecolor{udeLight}{RGB}{223,228,242}

\definecolor{coolblack}{rgb}{0.0, 0.18, 0.39}
\definecolor{sinopia}{rgb}{0.8, 0.25, 0.04}
\definecolor{ferrarired}{rgb}{1.0, 0.11, 0.0}

\definecolor{olive}{rgb}{0.5, 0.5, 0.0}

\newenvironment{definition}[1][Definition]{\begin{trivlist}
\item[\hskip \labelsep {\bfseries #1}]}{\end{trivlist}}

\definecolor{darkmidnightblue}{rgb}{0.0, 0.2, 0.4}

\usepackage{hyperref}
\hypersetup{
    colorlinks,
		pdfborder={0 0 0},
    linkcolor={ude},
    citecolor={ude},
    urlcolor={ude}
}

\newcommand{\figwidth}{0.95\columnwidth}

\newcommand{\E}[1][]{\ensuremath{{\rm{E}}\!\left[#1\right]}}
\newcommand{\Var}[1][]{\ensuremath{{\rm{Var}}\!\left[#1\right]}}

\DeclareFixedFont{\auacc}{OT1}{phv}{m}{n}{10} 
\DeclareFixedFont{\aunam}{OT1}{phv}{m}{n}{12}

\npstyleenglish

\usepackage{lipsum}

\newcommand{\bino}[1]{\ensuremath{\mathrm{Bino}(#1)}}

\usepackage{amsthm}

\definecolor{MatlabCellColour}{RGB}{252,251,220}
\usepackage[framed,numbered,autolinebreaks,useliterate]{mcode}

\newcommand{\binocase}{\textbf{Binomial distribution.}\xspace}
\newcommand{\lowvarcase}{\textbf{Low variance scenario.}\xspace}

\usepackage[small]{caption}

\begin{document}


\title{Confidence Interval Estimators for MOS Values }

\author{
 \IEEEauthorblockN{Tobias Ho{\ss}feld$^1$, Poul E. Heegaard$^2$, Mart{\'\i}n Varela$^3$,  Lea Skorin-Kapov$^{4}$}
 \IEEEauthorblockA{$^1$Chair of Communication Networks, University of W\"urzburg, Germany\\
$^2$NTNU, Norwegian University of Science and Technology, Trondheim, Norway\\
 $^3$callstats.io, Helsinki, Finland \\
 $^4$Faculty of Electrical Engineering and Computing, University of Zagreb, Croatia}
Email: $^1$tobias.hossfeld@uni-wuerzburg.de, $^2$poul.heegaard@ntnu.no, $^3$martin@callstats.io, $^4$lea.skorin-kapov@fer.hr
}

\maketitle

\begin{abstract}
  For the quantification of QoE, subjects often provide individual rating scores
  on certain rating scales which are then aggregated into Mean Opinion Scores
  (MOS). From the observed sample data, the expected value is to be
  estimated. While the sample average only provides a point estimator,
  confidence intervals (CI) are an interval estimate which contains the desired
  expected value with a given confidence level.
%
  In subjective studies, the number of subjects performing the test is typically small, especially in lab environments.
	The used rating scales are bounded and often discrete like the 5-point ACR rating scale.
  Therefore, we review statistical approaches in the literature for their
  applicability in the QoE domain for MOS interval estimation (instead of having only a point estimator, which is the MOS). 
	We provide a conservative estimator  based on the SOS hypothesis and binomial distributions and compare its
  performance (CI width, outlier ratio of CI violating the rating scale bounds)
  and coverage probability with well known CI estimators. 
	We show that the provided CI estimator works very well in practice for
  MOS interval estimators, while the commonly used studentized CIs suffer from a
  positive outlier ratio, i.e., CIs beyond the bounds of the rating scale. As an
  alternative, bootstrapping, i.e., random sampling of the subjective ratings
  with replacement, is an efficient CI estimator leading to typically smaller
  CIs, but lower coverage than the proposed estimator.	
\end{abstract}


\section{Introduction}

Quality of Experience (QoE) research commonly relies on the collection of
subjective ratings from a chosen panel of users to quantify various QoE
dimensions (also referred to as QoE features \cite{QoEWP}), e.g., related to
perceived audio/visual quality, perceived usability, or overall perceived
quality. While various rating scales have been used in both the user experience
(UX) and QoE research fields, the results of subjective studies reported by the
QoE community have to a large extent relied on the use of a standardized 5-point
Absolute Category Rating (ACR) scale to calculate Mean Opinion Score (MOS) values.
While it has been argued that researchers should go beyond the MOS in their
studies~\cite{hossfeld2016qoe} 
in order to consider
different applications and user diversity, MOS estimates remain a staple of the
QoE literature. 

In this context, the statistical analysis of subjective study results,
subsequently used to derive QoE estimation models~\cite{ITU_P1401}, relies on
the estimation of confidence intervals (CIs) to quantify the significance of MOS
values per test condition. Challenges arise in dealing with uncertainties
resulting from problems such as ordering effects and subject
biases~\cite{ITU_P1401,janowski2014_QoMEX}. Such statistical uncertainties are
expressed in terms of CIs. Given the nature of conducting QoE studies, two main
issues arise. Firstly, rating scales used in quantitative QoE evaluation are
bounded at both ends. Therefore, the individual rating scores $Y$ of a subject
are limited. 
However, for the calculation of CIs, normal distributions (due to central limit
theorem) or Student's t-distribution are used, which are unbounded.

Secondly, due to the inherent complexity of running subjective studies,
resulting in a compromise between a large number of test conditions and
participant fatigue, the number $n$ of test subjects taking part in a study is
generally small, in particular when running tests in a lab environment. We note
that while methods such as crowdsourcing may be utilized to obtain a much larger
population sample, in many cases the specifics of the study call for a
controlled lab environment. As an example, and bearing in mind that the number
of required participants clearly depends on the test design, number of test
conditions, and target population, the ITU-T recommends a minimum of 24 subjects
(controlled environment) or 35 subjects (public environment) for subjective
assessment of audiovisual quality \cite{ITU_P913}. ITU-T Recom. P.1401 further
states that if fewer than 30 samples are used, the normal distribution starts to
become distorted and calculation of CIs based on normality assumptions are no
longer valid. In cases with fewer than 30 samples, P.1401 advocates the
use of the Student t-distribution when calculating CIs.

Given the aforementioned issues, we highlight that commonly used CI estimators
do not work properly for small sample sizes, as the normal distribution
assumption may not be valid, and that they violate the bounds of the
rating scale. 
%
%
In this paper we review statistical approaches in the literature for their
application in the QoE domain for MOS interval estimation (instead of having
only a point estimator, which is the MOS). Due to space restrictions, we
consider only discrete rating scales, and test the CI estimators in terms of
efficiency (CI width), coverage (how many CIs overlap the true mean value), and
outlier ratio.
%

The remainder of this paper is organized as follows.
Section~\ref{sec:background} provides the background on
CIs 
such as the central limit theorem, used to derive CI estimators.
Section~\ref{sec:estimators} considers common estimators for the MOS and
introduces some estimators based on binomal distributions that are suitable for
MOS CI estimation. It also discusses other non-commonly used methods in the QoE
community, such as simultaneous CI and bayesian approaches for multinomial
distributions, as well as bootstrapping CI. Section~\ref{sec:results} defines
various scenarios for evaluating the performance of the estimators in terms of
coverage, outlier ratio, and CI width. Section~\ref{sec:conclusions} concludes
this work and gives some recommendations on CI estimators for MOS values in
practice.

\section{Background}\label{sec:background}
For the sake of readability, we briefly state the definitions and theorems used
to obtain an interval estimate, denoted {\em confidence intervals} in the
following. Table \ref{tab:notion} provides a summary of the notation used
throughout the paper.


\begin{definition}[Confidence interval: ] 
  Let $X$ be a random sample from a probability distribution with statistical
  parameter~$\theta$, which is a quantity to be estimated. The confidence
  interval $[\theta_0,\theta_1)$, is obtained by
\begin{equation}
P(\theta_0\leqslant\theta<\theta_1)=1-\alpha,\ 0<\alpha<1
\label{eq:ci}
\end{equation}
where $(1-\alpha)$ is the confidence coefficient (or \emph{degree of
  confidence}). The confidence interval contains the statistical
parameter~$\theta$ with probability $1-\alpha$.
\end{definition}

\begin{definition}[Central Limit Theorem (CLT): ]
Let $X$ be a random sample 
of size $n$ ($X=\{X_1,X_2,\ldots,X_n\}$) taken from a population with expected
value $\mathrm{E}(X_i)=\mu$ and variance $\mathrm{Var}(X_i)=\sigma^2<\infty$, $\
i=1,2,\ldots,n$, then the sample mean $\hat{X}$ asymptotically follows a normal
distribution with expected value $\mu$ and variance $\sigma^2/n$ as
$n\to\infty$.
\begin{equation}
\hat{X}\mathop{\sim}_{n\to\infty}\mathcal{N}\left(\mu,\frac{\sigma^2}{n}\right)
\label{eq:clt}
\end{equation}
\end{definition}


%

\begin{definition}[Confidence intervals for sample mean: ] 
  The confidence interval for the sample mean $\hat{X}$, with $E[X] = \theta$
  and standard error of the sample mean $S/\sqrt{n}$ according to CLT, can be
  obtained by 
\begin{equation}
\hat{X} \pm z_{\alpha/2} \cdot \frac{S}{\sqrt{n}}
\label{eq:ciMean}
\end{equation}
where $z_{\alpha/2}$ is the $\alpha/2$-quantile of the standard normal distribution $N(0,1)$.  $S^2$ estimates the unknown variance $\sigma^2$.
\end{definition}

This assumes that the sample size $n$ is large, and that the sampling
distribution is symmetric, which is not always the case. In the following, we
detail how to establish a confidence interval in the case of a sampling
distribution whose density function is symmetric or non-symmetric around the
mean.

Note that the variance of a sample mean, $\Var[\hat{X}] = \Var[\frac{1}{n}
\sum_{i=1}^n X_i] = \Var[X]/n$, where $\Var[X]$ is the variance of the sample
$X$. This implies that when $n\to \infty$ then $\Var[\hat{X}]\to 0$ while
$\Var[X] \to \sigma^2$.

%
%
%
%

\begin{table}%
\caption{Notation and variables used in the paper.}
\label{tab:notion}
\begin{tabularx}{\columnwidth}{lX}
\toprule
Variable             & Description \\
\midrule
$Y_x$                & random variable of user ratings for test condition $x$\\
$k$                  & users rate on a discrete $k$-point rating scale from $1,\dots,k$\\
$n$                  & number of users rating the test condition \\
$m$                  & number of test conditions (TC) \\
$r$                  & number of simulation runs \\
$y_{u,x,i}$          & sampled user rating for user $u$, TC $x$ and simulation run $i$\\
$\hat{Y}_{x,i}$ & MOS, i.e. sample mean over user ratings, for TC $x$ and run $i$ \\
$\gamma$             & confidence level \\
$\alpha$             & significane level, e.g. $\alpha=0.05$; it is $\alpha=1-\gamma$ \\
\bottomrule
\end{tabularx}
\end{table}
\section{Confidence Interval Estimators for MOS}\label{sec:estimators}

\subsection{Problem Formulation}
We assume we have a discrete rating scale with $k$ rating items, leading to a
multinominal distribution, which is a generalization of the binomial
distribution. For a certain test condition, $n$ users rate the quality on a
discrete $k$-point rating scale, e.g., $k=5$ for the commonly used 5-point ACR
scale. Each scale item is selected with probability $p_i$ for $i=1,\dots,k$;
$\sum_{i=1}^k p_i = 1$.

The $n$ users rate quality as one of the $k$ categories. Samples
$(n_1,\dots,n_k)$ indicate the number of ratings obtained per category, with
$\sum_{i=1}^k n_i = n$ (i.e., each user has provided one rating). With each
category having a fixed probability $p_i$, the multinomial distribution gives
the probability of any particular combination of numbers $n_i$ of successes for
the various categories (under the condition $n_k = n - \sum_{i=1}^{k-1} n_i$)

\begin{equation}
P(N_1=n_1,\dots,N_k=n_k) = \frac{n!}{n_1! \cdots n_k!} p_1^{n_1}\cdots p_k^{n_k}
\label{eq:multinominal}
\end{equation}

In QoE tests, we are interested in the rating of an arbitrary user. 
The marginal distribution (when $n=1$) with $p_i$ estimates the the expected rating $\E[Y]$ by the sample mean $\hat{Y}$ (aka MOS), 
assuming a linear rating scale.
\begin{equation}
	\hat{Y} = \frac{1}{k} \sum_{i=1}^k i p_i
\label{eq:unknownMOS}
\end{equation}

We denote $Y$ as a random variable of the rating of the users. We observe a
sample $Y_1, \dots, Y_n$ with $Y_i \in \{1,\dots,k\}$. As previously stated, in
subjective QoE tests, the number $n$ of users is typically not very high. From
the samples $(n_1,\dots,n_k)$, the MOS and CI can be estimated.
However, given the use of a bounded rating scale and small sample size, existing
estimators of CI do not follow the CLT and might be asymmetric around the sample
mean, and will potentially violate the bounds of the rating scale, i.e.,
$\theta_0 < 1$ and/or $\theta_1 > k$.

\subsection{Regular Normal and Student's t-distribution}
The most common way of constructing a CI from a set of samples, $X=\{ X_1,
\cdots ,X_n\}$, is to apply the CLT. When the variance of $X$ is not known, then
the quantile $t_{\alpha/2, n-1}$ must be taken from a Student's t-distribution
with confidence level $1-\alpha$ and $n-1$ degrees of freedom, unless the number
of samples are sufficiently large ($n>30$ according to ITU-T recommendation
P.1401). Then the quantiles in the Student's t-distribution and standard Normal
distributions are approximately the same.

The CI for both Student's t-distribution and Normal distribution is estimated by use of~\eqref{eq:ciMean}, the only difference is the quantiles.



Observe; truncating the upper and lower bounds, i.e., $\theta^*_0 = \max(1, \theta_0)$ and/or $\theta^*_1 = \min(k, \theta_1)$ is not correct.

\subsection{Simultaneous CIs for Multinomial Distribution}
A complementary approach is to consider the multinomial proportions $p_i$ of user ratings on the scale for item $i$ and then to derive exact confidence coefficients of simultaneous CI for those multinomial proportions. A method for computing the CIs for functions of the multinomial proportions is proposed in \cite{jin2013computing} which can be directly applied to the computation of the MOS, see ~Eq.(\ref{eq:unknownMOS}). There are $n_i$ user ratings for category $i$ and $\chi_{1-\alpha/k}$ is the quantile of the $\chi^2$-distribution with one degree of freedom considering $k$ simultaneous CIs. The MOS is $\hat{Y}=\sum_{i=1}^k i \frac{n_i}{n}$.

\begin{equation}
\sum_{i=1}^k i \frac{n_i}{n} \pm \sqrt{\frac{\chi_{1-\alpha/k}}{n} \left(\sum_{i=1}^k i^2 \frac{n_i}{n}\right)- \left( \sum_{i=1}^k i \frac{n_i}{n} \right)^2 }
\label{eq:simCI}
\end{equation}

\newcommand{\mk}{{k_0}}

\subsection{Using Binomial Proportions for Discrete Rating Scales}
The shifted binomial distribution can be used as an upper bound distribution for user rating distributions when users rate on a $k$-point rating scale ($1,\dots,k$). The binomial distribution leads to high standard deviations in QoE tests \cite{hossfeld2011sos} and follows exactly the SOS hypothesis with parameter $a=1/\mk$ with $\mk=k-1$. 

Let us consider $n$ users. Assume the user ratings follow a shifted binomial
distribution, $Y_i \sim \bino{\mk,p}+1$. Then, the sum of the user ratings
follows also a binomial distribution. 
	\begin{equation}
	Y=\sum_{i=1}^n Y_i \sim \bino{\sum_{i=1}^n \mk,p}+1=\bino{n \cdot \mk,p}+1
	\end{equation}
  and then $\hat{Y} = \frac{1}{n} \sum_{i=1}^n Y_i \sim \bino{\frac{1}{n}
    \sum_{i=1}^n \mk,p}+1=\bino{\cdot \mk,p}+1$. Due to differences among users,
  it may be $p_i \neq p_j$ for users $i$ and $j$. The binomial sum variance
  inequality can be used to derive an upper bound. Let us consider
  $Y=\sum_{i=1}^n Y_i$, which does not follow a binomial distribution. We
  define $Z\sim \bino{n\cdot \mk,\bar{p}}+1$ with $\bar{p}=\frac{1}{n}
  \sum_{i=1}^n p_i$. As a result of the binomial sum variance inequality we
  observe that the variance of $Z$ is an upper bound for QoE tests.
	\begin{equation}
	\Var[Y] < \Var[Z]
	\end{equation}

		%
Hence, we may use $\hat{Z}$ instead of $\hat{Y}$ to derive conservative CIs for the MOS based on the CI $[\hat{p}_0;\hat{p}_1]$ for the unknown $p$.
\begin{equation}
[\hat{Z}_0,\hat{Z}_1] = [\hat{p}_0,\hat{p}_1] \cdot (k-1) + 1
\label{eq:mos:bino}
\end{equation}

CI estimation for binomial distributions has drawn attention in the literature
and several suggestions have been provided. A few  works 
compare the CI estimators for binomial proportions
\cite{pires2008interval,vollset1993confidence,brown2001interval,newcombe2012confidence}.
For example, \cite{brown2001interval} suggests using Wilson interval and Jeffreys prior
interval for small $n$.
The normal theory approximation of a confidence interval for a proportion is
known as the Wald interval, which is however not recommended
\cite{agresti1998approximate}. For readability, we write $z=z_{\alpha/2}$ for
the $\alpha/2$-quantile of the standard normal distribution.
\subsubsection{Wald interval employing normal approximation} From the MOS $\hat{Y}$ we obtain $\hat{p}=\frac{\hat{Y}-1}{k-1}$. The standard deviation is $S=\sqrt{\hat{p}(1-\hat{p})}$. The CI for the MOS is as follows.
\begin{equation}
			(\hat{p} \pm z \frac{S}{\sqrt{n}})\cdot (k-1) + 1	\quad \Leftrightarrow \quad \hat{Y}\pm z \frac{S}{\sqrt{n}}  (k-1)
	\label{eq:wald}
	\end{equation}

\subsubsection{Wilson score interval with continuity correction} For the Wilson interval, a continuity correction is proposed which aligns the minimum coverage probability, rather than the average probability, with the nominal value.
\begin{align}
d &= 1+z \sqrt{(z^2-\frac{1}{n\mk} + 4n\mk\hat{p}(1-\hat{p})+(4\hat{p}-2))} \\
\hat{Y}_0 &= \max\left(1, \mk \frac{(2n\mk\hat{p} + z^2 - d }{(2(n\mk+z^2)} +1 \right)\\
\hat{Y}_1 &= \min\left(k, \mk \frac{(2n\mk\hat{p} + z^2 + d }{(2(n\mk+z^2)} +1 \right)
\end{align}

\subsubsection{Clopper-Pearson} It is the central exact interval \cite{clopper1934use} and we use the implementation based on the beta distribution with parameters $c$ and $d$ \cite{agresti1998approximate}. The parameter $c$ quantifies the number of `successes' of the corresponding binomial proportion, i.e. $c=\sum_{i=1}^n{(y_i-1)}$ for user ratings $y_i$, and $d=n(k-1)-c+1$. The $q$-quantile of the beta distribution is denoted by $\beta_q(c,d)$.
\begin{align}
\hat{Y}_0 &= \max\left(1, b_{\alpha/z}(c,d) \cdot (k-1)+1 \right)\\
\hat{Y}_1 &= \min\left(k, b_{1-\alpha/z}(c,d) \cdot (k-1)+1 \right)
\end{align}

\subsubsection{Jeffreys Interval} A Bayesian approach for binomial proportions is Jeffreys interval which is an exact Bayesian credibility interval and guarantees a mean coverage probability of $\gamma$ under the specified prior distribution. \cite{brown2001interval} have chosen the Jeffreys prior \cite{jeffreys1998theory}. Although it follows a different paradigm, it has also good frequentist properties and looks similar to Clopper-Pearson. The calculation also uses the number of successes $c$ as defined above and the quantiles of the beta distribution.
\begin{align}
\hat{Y}_0 &= \begin{cases}
							  b_{\alpha/2}(c+\frac{1}{2},d- \frac{1}{2}) \cdot (k-1)+1 & \\
								0 \, \text{ if } c=0
							\end{cases} \\
\hat{Y}_1 &= \begin{cases}
							  b_{1-\alpha/2}(c+\frac{1}{2},d- \frac{1}{2}) \cdot (k-1)+1 & \\
								k \, \text{ if } c=n(k-1)
							\end{cases} 
\end{align}

\subsection{Bootstrap Confidence Intervals}
The non-parametric bootstrap method as introduced by
Efron~\cite{efron1992bootstrap} uses solely the empirical distribution of the
observed sample. Simulations from the empirical distribution lead to many
observations of various MOS estimators $\hat{Y_r}$ for each simulation run $r$.
As a result, a distribution of mean values is observed and the CIs can be
directly obtained based on Eq.~\eqref{eq:ci}. We use Matlab's implementation
of the `bias corrected and accelerated percentile' method to cope with the
skewness of the observed distribution, cf.~\cite{efron1992bootstrap}.

\section{Numerical Results}\label{sec:results}
For evaluating the estimators' performance, we consider different scenarios in
which the user ratings for a test condition are sampled from a known
distribution. The commonly used 5-point ACR scale is considered. We
investigate 
two different scenarios:
\begin{inparaenum}[(1)]
	\item binomial distribution as an upper bound in terms of variance for QoE
    tests, 
	\item low variance, where users only rate $2,3,4$ and avoid the rating scale edges.
\end{inparaenum}
The performance is then evaluated with several metrics: the coverage of
the CIs, the width of the CIs, and the outlier ratio.

\subsection{Scenarios for Performance Evaluation}
We consider a $k$-point rating scale. For a certain test
condition $x$, the user ratings $Y_x$ follow a certain discrete distribution,
with $p_i=P(Y_x=i)$ for $i \in{1, ..., k}$.
User ratings $Y_{u,x,i}$ are sampled for test condition $x$ for the users $u \in \{1,\dots,n\}$, from the 
distribution $F_{Y_x}$. 
The simulations are repeated $r$-times to get statistically significant results in the evaluation. The index $r \in \{1,\dots,r\}$ represents the $r$-th simulation run. We use $r=200$ repetitions.
%
For the evaluation, we consider $m=101$ test conditions 
with the known mean value, i.e., the \emph{expected value}, $\E[Y_x] = \mu_x $ for $x \in\{1,\dots,m\}$. It is $\mu_x = \frac{x-1}{m}(H-L)+L$ with $H \leq k$ and $L \geq 1$ indicating the maximum and minimum possible user rating $Y_x$, respectively. 

\subsubsection{Binomially Distributed User Ratings}
  This scenario represents a high variance of user ratings which is also observed in real QoE tests. The user rating diversity for any QoE experiment can be quantified in terms of the SOS parameter $a$ which is defined in \cite{hossfeld2011sos}. For example, \cite{hossfeld2016qoe} measured $a=0.27$ for the results of a web QoE study. This was among the highest SOS parameters observed for different QoE studies and applications such as video streaming, VoIP, and image QoE. The results of gaming QoE studies have shown a similarly high SOS parameter. The binomial distribution leads to an SOS parameter of $a=0.25$ and is therefore appropriate as a realistic scenario for high variances. 	
	
	\begin{equation}
	Y_x \sim \bino{k-1,p}+1
	\end{equation}

with MOS $\E[Y_x]=p \cdot (k-1)+1$ and $\Var[Y_x]=(k-1)p(1-p)$. Hence, $p=\frac{x-1}{k-1}$.

\subsubsection{Low Variance}
Next, we consider a scenario with low variances. In that case, users are not using the edge of the rating scale and only rate $2,\dots,k-1$. This can be realized with a shifted binomial distribution. 

	\begin{equation}
	Y_x \sim \bino{k-3,p}+2	
	\end{equation}
	 with $\E[Y_x]=p (k-3)+2$ and $\Var[Y_x]=(k-3)p(1-p)$. Then $p=2 \frac{x-1}{k-1}-\frac{1}{2}$.
	The SOS parameter is numerically derived \cite{hossfeld2016qoe} and found to be $a=0.084$. 


\subsection{Metrics for Evaluating the Performance of the Estimators}
According to the distribution defined in a given scenario, we generate $n$ samples (i.e., user ratings) for $m$ test conditions and repeat the simulation $r$ times. The user rating $y_{u,x,i}$ indicates the user rating of user $u$, test condition $x$, in run $i$. 

For each test condition $x$ and each run $i$, the MOS is derived by averaging over the $n$ sampled subjects' ratings.

\begin{equation}
\hat{Y}_{x,i} = \frac{1}{n}\sum_{u=1}^n y_{u,x,i}
\label{eq:sampleMOS}
\end{equation}

The CI estimator does not know the underlying distribution $Y_x$ or
the 
expected values $\mu_x$. We investigate the performance of the CI estimator 
with the following metrics.

\subsubsection{Coverage} For a certain confidence interval derived from the
samples of all $n$ users for test condition $x$ in run $i$, we can check whether
the 
expected value $\mu_x$ is contained in the confidence interval
$[\theta_L;\theta_U]$.
\begin{equation}
C_{x,i} = 
\begin{cases}
	1 & \text{if } \theta_L \leq \mu_x \leq \theta_U \\
	0 & \text{otherwise}
\end{cases}
\end{equation}

Then, the coverage of the CI estimator for test condition $x$ is the average over all $r$ simulation runs, i.e., the probability that the CI contains the expected value. 
The marginal distribution of $C_{x,i}$ for a fixed test condition $x$, gives the 
\emph{test condition perspective} and will be defined accordingly for the CI width and the outlier ratio.
\begin{equation}
\hat{C}_x = \frac{1}{r} \sum_{i=1}^r C_{x,i}
\label{eq:coverageCx}
\end{equation}


The marginal distribution of $C_{x,i}$ for a single QoE study, $i$, gives the 
\emph{QoE study perspective}.
\begin{equation}
\hat{C}_i = \frac{1}{m} \sum_{x=1}^m C_{x,i}
\label{eq:coverageCi}
\end{equation}
Please note that the overall average over all studies and test condition
$\hat{C}$ is obtained either by averaging over $\hat{C_x}$ or $\hat{C_i}$ . 
%
\begin{equation}
\hat{C} = \frac{1}{m} \sum_{x=1}^m \hat{C}_x =  \frac{1}{r} \sum_{i=1}^r \hat{C}_i
\label{eq:averageCoverage}
\end{equation}

\subsubsection{Outlier Ratio}
For test condition $x$ and study $i$, we estimate the probability that the 
confidence interval $[\theta_0;\theta_1]$ is outside the bounds of the rating scale $[1,k]$.
\begin{equation}
O_{x,i} = 
\begin{cases}
	1 & \text{if } \theta_0<1 \text{ or } \theta_1>k \\
	0 & \text{otherwise}
\end{cases}
\end{equation}
Then, we define the outlier ratio from the test condition perspective and the QoE study perspective, respectively.
\begin{equation}
\hat{O}_x = \frac{1}{r} \sum_{i=1}^r O_{x,i} \, , \quad \hat{O}_i = \frac{1}{m} \sum_{x=1}^m O_{x,i} \, .
\label{eq:outlier}
\end{equation}
%

\subsubsection{CI Width} 
Finally, the width $\hat{W}_x$ and $\hat{W}_i$ of the confidence intervals is considered from the test condition perspective and the QoE study perspective, respectively. Thereby, the confidence intervals are averages over all runs and over all test conditions, respectively.
\begin{equation}
\hat{W}_x =  \frac{1}{r} \sum_{i=1}^r W_{x,i} \, , \quad \hat{W}_i = \frac{1}{m} \sum_{x=1}^m W_{x,i} \, .
\label{eq:ciwidth}
\end{equation}
Please note that the average over $\hat{W}_x$ and the average over $\hat{W}_i$ are identical.
\begin{equation}
\hat{W} = \frac{1}{m} \sum_{x=1}^m \hat{W}_x =  \frac{1}{r} \sum_{i=1}^r \hat{W}_i
\label{eq:averageCoverage}
\end{equation}

\subsection{Scenario with Binomially Distributed Ratings}


\begin{figure}%
\centering%
\includegraphics[width=\figwidth]{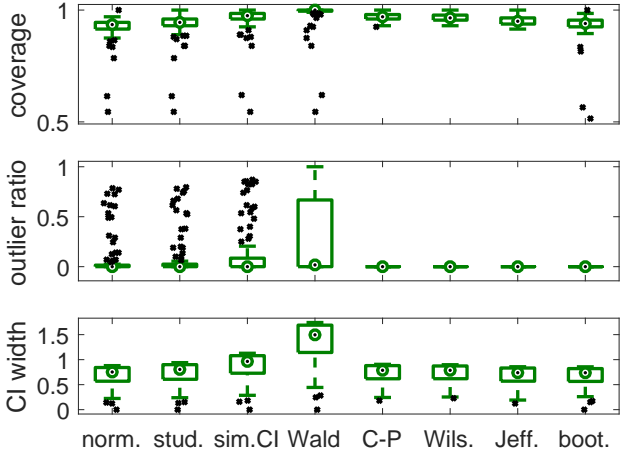}%
\caption{\binocase The \emph{test condition perspective} considers the performance measures $\hat{M}_x$. We observe that for some estimators (norm., stud., sim.CI, Wald) there are several test conditions with bad properties (low coverage, high outlier ratio). The corresponding numbers are provided in Table~\ref{tab:binoStudyBothTables}. Except for the Wald estimator, the binomial proportion estimators (C-P, Wils., Jeff.) work much better. Bootstrapping also leads to good results, but suffers from coverage outliers. 
}%
\label{fig:binoPerspectiveSingleInterval}%
\end{figure}

\begin{figure}%
\centering%
\includegraphics[width=\figwidth]{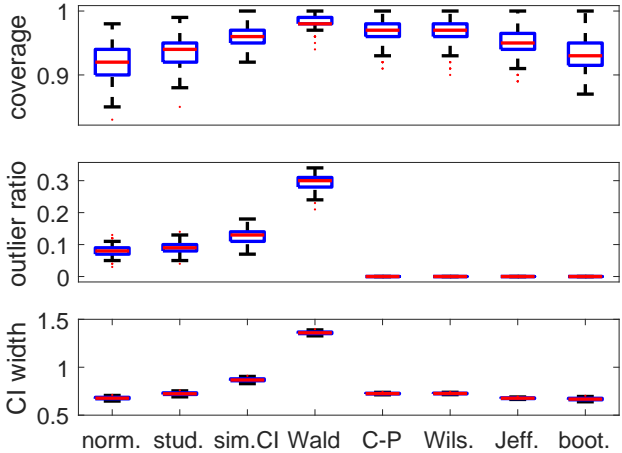}%
\caption{\binocase The \emph{QoE study perspective} focuses on the performance measure $\hat{M}_i$. Hence, the performance (coverage, outlier ratio, CI width) is averaged over all test conditions within a single run. The boxplot summarizes then those average results $\hat{M}_i$ over all $r$ runs. Concrete numbers are provided in Table~\ref{tab:binoStudyBothTables}.
}%
\label{fig:binoPerspectiveQoEStudy}%
\end{figure}

\newcommand{\hi}[1]{\textcolor{red}{#1}}
\definecolor{bronze}{rgb}{0.8, 0.5, 0.2}
\newcommand{\yi}[1]{\textcolor{bronze}{#1}}

\begin{table}\centering
\caption{ The performance metrics are averaged and differentiated for coverage from \emph{test condition perspective} ($\hat{C}_i$) and a \emph{QoE study perspective} ($\hat{C}_x$) for the three scenarios. Minimum coverage is denoted by $\hat{C}_{x|i}^m$ and coverage outliers in the boxplot by  $\hat{C}_{x|i}^o$.
}
\label{tab:binoStudyBothTables}  
\begin{tabular}{cccccccc}
\toprule                 
\emph{Binomial} & $\hat{C}$ & $\hat{C}_x^o$ & $\hat{C}_x^m$ & $\hat{C}_i^o$ & $\hat{C}_i^m$ & $\hat{O}$ & $\hat{W}$ \\
\midrule
norm. & 0.92 & 0.08 & \hi{0.55} & 0.01 & 0.83 & \hi{0.08} & 0.68 \\
stud. & 0.93 & 0.09 & \hi{0.55} & 0.01 & 0.85 & \hi{0.09} & 0.72 \\
sim.CI & 0.96 & 0.08 & \hi{0.55} & 0.00 & 0.92 & \hi{0.13} & 0.87 \\
Wald & 0.98 & 0.14 & \hi{0.55} & 0.04 & 0.94 & \hi{0.30} & \hi{1.36} \\
C-P & 0.97 & 0.01 & 0.93 & 0.03 & 0.91 & 0.00 & 0.72 \\
Wils. & 0.97 & 0.00 & 0.93 & 0.04 & 0.90 & 0.00 & 0.73 \\
Jeff. & 0.95 & 0.00 & 0.92 & 0.04 & 0.89 & 0.00 & 0.68 \\
boot. & 0.93 & 0.05 & \hi{0.52} & 0.00 & 0.87 & 0.00 & 0.67 \\      
\toprule
\emph{Low. var.} & $\hat{C}$ & $\hat{C}_x^o$ & $\hat{C}_x^m$ & $\hat{C}_i^o$ & $\hat{C}_i^m$ & $\hat{O}$ & $\hat{W}$ \\
\midrule
norm. & 0.90 & 0.10 & \hi{0.28} & 0.00 & 0.82 & 0.00 & 0.48 \\                                   
stud. & 0.91 & 0.09 & \hi{0.28} & 0.00 & 0.83 & 0.00 & 0.51 \\                                  
sim.CI & 0.93 & 0.10 & \hi{0.28} & 0.01 & 0.87 & 0.00 & 0.61 \\                                   
Wald & 1.00 & 0.00 & 1.00 & 0.00 & 1.00 & 0.00 & \hi{1.67} \\                                      
C-P & 1.00 & 0.23 & 0.98 & 0.14 & 0.99 & 0.00 & \yi{0.87} \\
Wils. & 1.00 & 0.24 & 0.98 & 0.16 & 0.99 & 0.00 & \yi{0.87} \\
Jeff. & 1.00 & 0.05 & 0.98 & 0.23 & 0.97 & 0.00 & \yi{0.82} \\ 
boot. & 0.91 & 0.11 & \hi{0.28} & 0.01 & 0.83 & 0.00 & 0.47 \\    
\bottomrule 
\end{tabular}
\end{table} 

Figures~\ref{fig:binoPerspectiveSingleInterval}
and~\ref{fig:binoPerspectiveQoEStudy} show the results for the binomial
distribution scenario for the TC and QoE study perspective, respectively. 

The boxplots shows the median within the box. The bottom and top of the box are the first and third quartiles.  
The upper and lower ends of the whiskers denotes the most extreme data point that is maximum and minimum 1.5 interquartile range (IQR) of the upper and lower quartile, respectively. Data outside 1.5 IQR are marked as outlier with a dot.
%

An overview on the performance measures is provided in
Table~\ref{tab:binoStudyBothTables}. The numerical results from the binomial
case show that Clopper-Pearson, Jeffreys and bootstrapping have a good
performance from the test condition and QoE study perspective. They have a good
coverage, do not suffer from outliers, and have small CI widths.
%
%
%

The proposed idea based on binomial proportion fails if the distribution has a
higher variance than a binomial distribution. Then, the coverage is poor; the
confidence intervals are too small, as only binomial variances are assumed, but
in reality we have higher variances. This is however very rare in
actual QoE studies. If the variances are higher, this is often an
indicator for hidden influence factors in the test setup or some other issues
\cite{hossfeld2011sos}. 

\subsection{Low Variance Scenario}
\begin{figure}%
\centering%
\includegraphics[width=\figwidth]{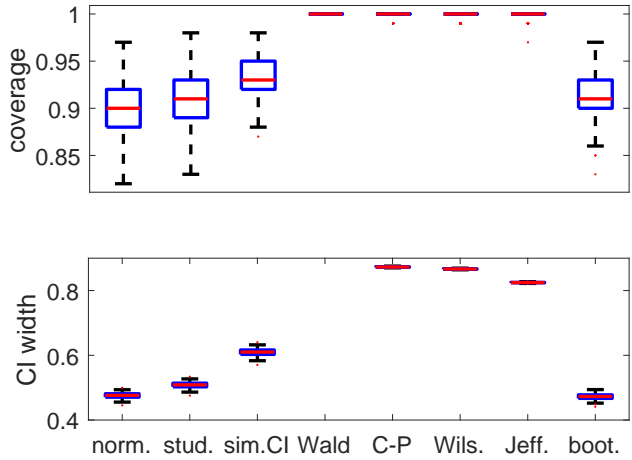}%
\caption{\lowvarcase For all estimators, the outlier ratio is zero. Wald interval average CI width is about 1.67. The proposed binomial based CBI estimators lead to higher CIs than normalized estimators, as the assumed binomial distribution has a higher variance. Thus, the estimators are conservative for low variances. 
}%
\label{fig:lowvarPerspectiveQoEStudy}%
\end{figure}

We only consider the QoE study perspective now which is provided in Figure~\ref{fig:lowvarPerspectiveQoEStudy}.
In case of low variances, the three identified estimators (Wilson, Clopper-Pearson, Jeffreys) still have a very good performance, and coverage is 100\%. However in that case, the CI width is larger than for the normalized or student-t estimators. The reason for this is that the proposed estimators assume a binomial distribution (i.e., a much larger variance) and necessarily overestimate the CIs. For all estimators, the outlier ratio is zero. Still normalized or student-t have some problems to cover certain TCs at the edge (see $\hat{C}^m_x$ or $\hat{C}^o_x$).

\begin{figure}%
\centering%
\includegraphics[width=\figwidth]{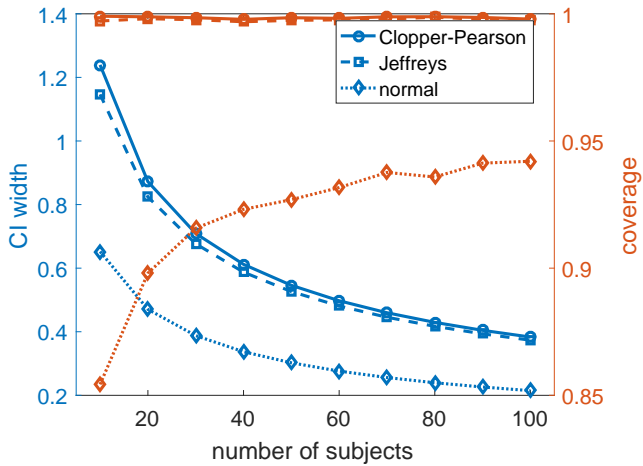}%
\caption{\lowvarcase The average coverage $\hat{C}$ and CI width $\hat{W}$ are considered depending on the number of subjects of the study. The outlier ratio is zero for the three considered estimators. 
}%
\label{fig:lowvarciWidthCoverageScatter2}%
\end{figure}

Figure~\ref{fig:lowvarciWidthCoverageScatter2} considers the average CI width and coverage when varying the number $n$ of subjects in the study. The most efficient way to decrease the CI width is to increase $n$. It is worth to note that the binomial proportions estimators show almost constant coverage in contrast to bootstrapping.

\section{Recommendations and Conclusions}\label{sec:conclusions}
Subjective QoE studies often involve a relatively small number of test participants. Moreover, used rating scales are commonly discrete and bounded at both ends, with study results reported in the form of MOS values and CIs derived for various test conditions to quantify the significance of MOS values. Given the importance of using efficient CI estimators  in the context of deriving QoE models, we evaluate several MOS CI estimators, and develop our own estimator based on binomial proportions. 
The numerical results indicate that the proposed idea based on binomial estimators is robust and conservative in practice.
Wilson, Clopper-Pearson, and Jeffreys lead to comparable results, with excellent coverage and outlier properties. 
However, very good coverage comes along with costs of having larger CI widths. The Wald interval performs poorly, unless $n$ is quite large, which is not commonly the case in QoE studies.
Standard confidence intervals based on normal and student-t distribution, as well as
simultaneous CIs for multinomial distributions, suffer from the CIs
exceeding the bounds of the rating scale. Bootstrapping has similar issues, i.e., some test conditions are not captured properly, but the outlier ratio is always zero due to sampling. 


In summary, for QoE tests characterized by a small sample size and the use of discrete bounded rating scales, the proposed binomial estimators (Clopper-Pearson, Wilson, Jeffreys) are conservative, but exact and recommended. For decreasing the CI widths, bootstrapping or standard CI may be used in case of low variance (when the SOS parameter $a<0.1$) at the cost of decreased coverage -- but the most effective way is to increase the number of subjects. If the SOS parameter is larger than for a binomial distribution ($a>\frac{1}{k-1}$), the results and test design should be checked, as there may be hidden influence factors in the study.
An implementation of the CI estimators and the recommended estimators based on the SOS parameter is available in Github \url{https://github.com/hossfeld}. 




\bibliographystyle{IEEEtran}
\bibliography{literature}

\end{document}